%% file: main.tex
\documentclass[12pt]{article}
\usepackage{amsmath}
\usepackage{bm}
\usepackage{graphicx}
\usepackage{enumerate}
\usepackage{natbib}
\usepackage{url}  
\usepackage{lmodern}    
\usepackage{ifthen}  
\usepackage{amsthm}
\usepackage{mathtools}
\usepackage{enumitem}
\usepackage{authblk} 
\usepackage{hyperref}
\usepackage[title,page]{appendix}
\hypersetup{colorlinks=true,linkcolor=blue,urlcolor=blue,citecolor=blue}
\usepackage{multirow}

\usepackage{amssymb}

\newcommand{\inner}[2]{\langle #1, #2 \rangle} 

\theoremstyle{plain}
\newtheorem{theorem}{Theorem} 

\theoremstyle{remark}

\DeclareMathOperator*{\argmin}{arg\,min}

\newcommand{\blind}{1}

\begin{document}

\def\spacingset#1{\renewcommand{\baselinestretch}%
{#1}\small\normalsize} \spacingset{1}

\renewcommand\Authands{ and }

\if1\blind
{
  \title{\bf  A Penalized Functional Linear Cox Regression Model for Spatially-defined Environmental Exposure with an Estimated Buffer Distance}
  \author[1]{Jooyoung Lee\thanks{Corresponding author. Email: jooylee@cau.ac.kr}}
  \author[8,9]{Zhibing He}
  \author[2,6,7]{Charlotte Roscoe}
  \author[5,6]{Peter James}
  \author[3]{Li Xu}
  \author[9,10]{Donna Spiegelman}
  \author[11]{David Zucker}
  \author[2,3,4]{Molin Wang}

  \affil[1]{Department of Applied Statistics, Chung-Ang University, Seoul Korea}
  \affil[2]{Channing Division of Network Medicine, Department of Medicine, Brigham and Women’s Hospital and Harvard Medical School, MA USA}
  \affil[3]{Department of Epidemiology, Harvard T.H. Chan School of Public Health, MA USA} 
  \affil[4]{Department of Biostatistics, Harvard T.H. Chan School of Public Health, MA USA} 
  \affil[5]{Division of Chronic Disease Research Across the Life Course (CoRAL), Department of Population Medicine, Harvard Medical School and Harvard Pilgrim Health Care Institute, MA USA}
  \affil[6]{Department of Environmental Health, Harvard T.H. Chan School of Public Health, MA USA}
  \affil[7]{Division of Population Sciences, Dana Farber Cancer Institute, MA USA}
  \affil[8]{Department of Biostatistics, Center for Methods on Implementation and Prevention Science, CT USA}
  \affil[9]{Department of Biostatistics, Yale School of Public Health, CT USA}
  \affil[10]{Center for Methods on Implementation and Prevention Science, Yale School of Public Health, CT USA}
  \affil[11]{Department of Statistics and Data Science, Hebrew University, Jerusalem Israel}
  \maketitle
} \fi



\if0\blind
{
  \bigskip
  \bigskip
  \bigskip
  \begin{center}
    {\LARGE\bf A Penalized Functional Linear Cox Regression Model for Spatially-defined Environmental Exposure with an Estimated Buffer Distance}
\end{center}
  \medskip
} \fi

\bigskip
\begin{abstract}

In environmental health research, it is of interest to understand the effect of the neighborhood environment on health. Researchers have shown a protective association between green space around a person's residential address and depression outcomes. In measuring exposure to green space, 
distance buffers are often used. However, buffer distances differ across studies. Typically, the buffer distance is determined by researchers \emph{a priori}. It is unclear how to identify an appropriate buffer distance for exposure assessment. To address geographic uncertainty problem for exposure assessment, we present a domain selection algorithm based on the penalized functional linear Cox regression model. The theoretical properties of our proposed method are studied and simulation studies are conducted to evaluate finite sample performances of our method. The proposed method is illustrated in a study of associations of green space exposure with depression and/or antidepressant use in the Nurses' Health Study. 
\end{abstract}

\noindent%
{\it Keywords: Functional linear Cox regression model; B-spline basis functions; Group bridge Lasso; neighborhood environment; buffer distance.}  
\vfill

\newpage
\spacingset{1.5} 

\input{sec1} 
\input{sec1_2} 
\input{sec2} 
\input{sec3} 
\input{sec4} 
\input{sec8} 
\input{sec9} 
\input{sec10} 

\pagebreak
\bibliographystyle{agsm}
\bibliography{ref}


\end{document}

%% file: sec1.tex
\section{Introduction}
\label{sec:intro}
Spatial uncertainty in environmental exposure assessment is a critically important issue in environmental epidemiology, as it can bias health effect estimates \citep{spiegelman2010approaches}.
In environmental exposure assessment, researchers often query spatial datasets using a geographic information system (GIS) to quantify the environment surrounding residential addresses, with the objective of assessing associations of environmental exposure and health across many residents. To assess spatial exposures, spatial data are typically quantified (e.g., spatially averaged) based on circular distance buffers surrounding addresses, which are created by specifying multiple distances (radii) and drawing concentric circles at the specified distances around the address points. Typically, the radii of circular buffers are selected by researchers \emph{a priori} based on hypothesized mechanistic pathways. For example, a larger buffer radius compared to a smaller buffer radius may be deemed appropriate to capture the environment relevant for physical activity, whereas the reverse may be hypothesized if the researcher aims to assess acoustic properties directly around participants’ homes. Selection of buffers \emph{a priori}, however, may lead to inconsistency in exposure assessment across epidemiological studies.  If we use a smaller buffer radius than the true buffer radius, we ignore the contribution of environmental exposures in the effect estimate outside the buffer radius and bias may occur since environmental exposures in the near distance are correlated. On the other hand, if we use a larger buffer radius than the correct buffer radius, then the variance of the effect estimate may increase \citep{baek2016distributed}.  In addition, the effect estimate may be null in the intermediate area, and the traditional approach does not work in this setting.  Therefore, spatial uncertainty in exposure assessment may hinder accurate inference of associations with health outcomes, as well as compromise comparability across studies. These concerns, in turn, may affect the translation of research findings into actionable environmental policy. 
\cite{kwan2012uncertain} defined this issue as the Uncertain Geographic Context Problem, or ``the spatial uncertainty in the actual areas that exert the contextual influences under study.'' In the context of environmental epidemiology, this can be interpreted as uncertainty in selecting the correct buffer surrounding residential addresses to observe the influence of an environmental exposure on a health outcome. Previous studies have used multiple discrete buffer radii for spatial environmental exposures and shown that associations between exposure and health measures are sensitive to different buffer radii \citep{james2014effects}. However, finding the limit beyond which the environment has no influence on health is difficult since the additional effect of exposure by increasing buffer radius cannot be estimated. 

In this paper, we measure environmental exposures in concentric rings with multiple radii, investigate the association between exposure on health outcomes as a function of radius, and identify a radial distance beyond which no association between exposure and health is observed. The environmental exposures can be statistically expressed using functional covariates as a function of radius within the framework of functional regression models. Then, finding areas in which environmental exposure is associated with the health outcome can be translated into the problem of identifying the domain of functional covariates in which they may affect the response in the functional regression models \citep{james2009functional}.  In other words, by solving the domain selection problem in the functional regression models with survival data, we can determine relevant areas for a particular exposure-health association.   

Functional linear regression models were originally developed by \cite{ramsay1997functional} to address the problem of estimating regression coefficients when covariates are functions. Functional linear Cox regression models have been recently proposed to address the relationship between functional covariates and survival outcomes. \cite{gellar2015cox} maximized the penalized partial likelihood using B-splines smooth functions and \cite{qu2016optimal} formulated the functional Cox regression model through a reproducing kernel Hilbert space.  \cite{kong2018flcrm} adopted the functional principle component analysis framework in the time-to-event settings. 
To our knowledge, however, the domain selection problem has not been addressed in functional linear Cox regression models. In functional linear regression models, a method for obtaining smooth and locally sparse estimators based on the penalized least squares method was developed by \cite{lin2017locally} to identify the null regions where the coefficient functions are zero; that is, the regions where no association of exposure and outcome is observed. Additionally, to find a truncation point in a functional linear regression model, \cite{hall2016truncated}  proposed a penalized least squares method with a penalty on a truncation point and \cite{guan2020estimating} adopted a nested group bridge penalty to estimate a truncation point, which can be used to identify the optimal/inflection point across exposures. While the extension of \cite{lin2017locally} to the Cox regression model can be implemented to find regions with null-effect, their method relies on the value of a pre-determined threshold to select variables based on the local quadratic approximation. The method of \cite{guan2020estimating} is restricted to the special case, in which the coefficient functions can be zero only in the tail region of the distribution, and uses a nested penalty structure. However, it is often inadequate to impose increasing penalties as the buffer distance increases in environmental exposure applications. 

In this paper, we develop methods for identifying the areas corresponding to non-null regions of coefficient functions in survival data analysis settings. We build on a combination of the functional Cox model literature and the sparse functional data analysis literature and use a more flexible method for allowing for regions where the coefficient function is equal to zero. We consider a penalized scalar-on-function linear Cox regression model to estimate the coefficient functions, as well as to select important domains. We use a B-spline basis to represent the coefficient function. We adopt a smoothness penalty to enforce smoothness of the coefficient function and a group bridge penalty for the selection of spline coefficients \citep{huang2009group, huang2014group}. Our aim is to address geographic uncertainty via data-informed identification of optimal exposure areas as well as to estimate the association of environmental exposure and survival outcomes within these non-null regions. In addition, by investigating how exposure-health association changes with the distance from the residence, it is possible to find an area where the strongest association is observed. Our proposed method is, in general, to identify non-null regions. However, it can be also applied to the special case where the objective is to identify a certain location after which there is no association by identifying the largest non-null location.

This article is organized as follows. In Section \ref{sec:motivate}, we describe the motivating data set used in our case study including the green space exposure and the Nurses’ Health Study. In Section \ref{sec:meth}, we present the proposed methods for functional linear Cox regression models with the smooth and group bridge penalties. In addition, we derive the theoretical properties of the proposed estimators. 
In Section \ref{sec:simu}, we examine the performance of our method using simulation studies. In Section \ref{sec:data}, we apply our method to an illustrative example to examine the association between greenspace exposure and depression incidence in the Nurses’ Health Study (NHS). We provide concluding remarks in Section \ref{sec:dis}. 

%% file: sec1_2.tex
\section{Motivating Data Set}\label{sec:motivate}

Depressive disorders have prevailed as one of the leading causes of disability, globally, for the last 30 years \citep{james2018global}. They affect more than 264 million people worldwide \citep{world2017depression}. Biological predisposition and psychosocial factors can contribute to depression onset \citep{world2017depression}; however, modifiable environmental exposures may also play a role in increasing risk for, or protecting against, depression \citep{van2019environmental}. Researchers have shown a protective association between urban greenspace surrounding the residential addresses and depression or depressive symptoms across multiple contexts \citep{sarkar2018residential, pun2018association, liu2019neighbourhood}. 

Our investigation is motivated by a study by \citet{banay2019greenness}, which showed that women in the Nurses’ Health Study (NHS) who lived in the most-green versus least green exposure quintile showed a lower risk of clinically-diagnosed depression and/or antidepressant use. The authors used two distance buffers (radii: 250m, 1250m) to assess greenness exposure, which were selected \emph{a priori}, and which, as the authors acknowledged, may introduce bias due to geographic uncertainty. Statistically significant associations between greenspace and depression were observed in the smaller buffer (250m), though not the larger buffer (1250m); however, the limit at which the exposure no longer exerts an effect on the outcome is unknown. In the present paper,  using higher resolution greenness data, we aim to identify the radial distance beyond which greenness exposure has no influence on depression incidence. 

The NHS is a prospective cohort study of U.S. female nurses living across 11 states. It was established in 1976, when a total of 121,701 nurses aged 30-55  were recruited. Nurses receive biennial questionnaires on lifestyle risk factors and health outcomes. In 2000,  depression diagnosis statuses and antidepressant prescription items were included on the questionnaire. The current analysis included all women who returned questionnaires in 2000 and had no diagnosis of depression before 2000. The 35,283 eligible participants contributed 296,436 person-years of follow-up, and 3,250 incident depression cases occurred between 2000 and 2010. 

\begin{figure} 
\centering
\includegraphics[scale=0.6]{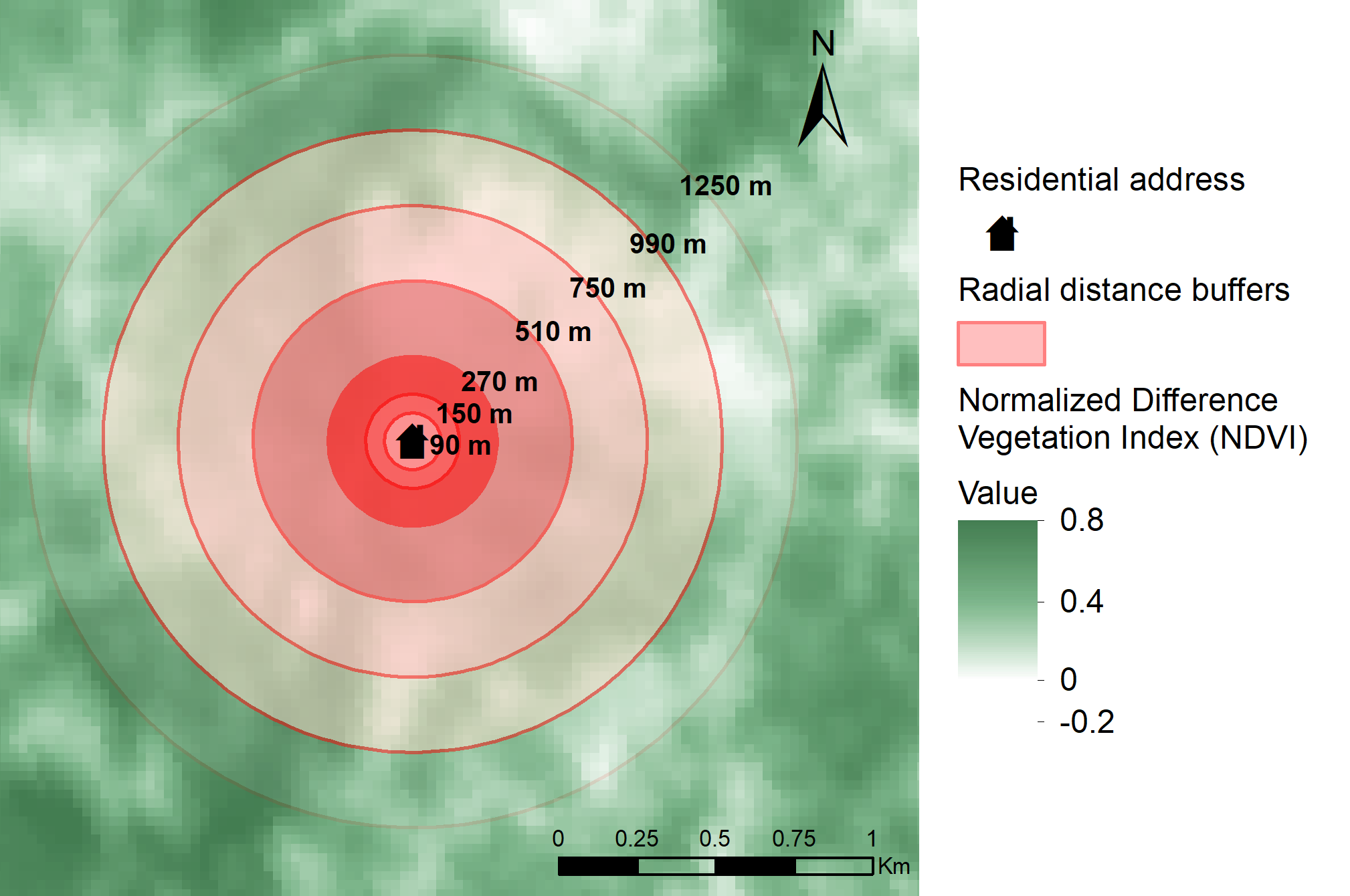}
\caption{An example of Normalized Difference Vegetation Index (NDVI) data surrounding an example residential address with multiple radial distance buffers.}
\label{Figure1}
\end{figure}
 
Normalized Difference Vegetation Index (NDVI) values were used to quantify greenness surrounding residential addresses of NHS participants. NDVI is a satellite-derived indicator of quantity of photosynthesizing vegetation; it is widely used as a marker for exposure to greenspaces in epidemiological studies \citep{rhew2011validation}. 
NDVI data derived from Landsat-7 satellite imagery were ascertained and processed via Google Earth Engine (GEE). We used the LandsatSimpleComposite algorithm to compute a Landsat Top of Atmosphere composite from 30m resolution raw scenes and selected the least-cloudy scenes. We computed the normalized difference between the near-infrared and red bands to calculate NDVI. From the 30m NDVI composite, we used the GEE reduceNeighborhood algorithm to create 30m resolution surfaces for the contiguous US in which each pixel represented the average of the NDVI values within a radial buffer surrounding that pixel in the input image (radii: 90m, 150m, 270m, 510m, 750m, 990m, 1230m, 1500m or 2100m) in 2000 (Figure \ref{Figure1}). Pixels were included in the calculation of average NDVI if 50\% of their surface was within the radial buffer. NDVI values range from 0 to 1, with higher values indicating greater cover of photosynthesizing vegetation. NDVI values below 0 indicate water and were set to 0. We spatially joined this NDVI dataset to the location of the residential addresses of NHS participants using the raster package (\url{https://rspatial.org/raster}) in R. 
 NDVI reaches its maximum and highest level of geographic variation during the summer. We therefore used the mean NDVI value of summer months (July 1 – September 30) in 2000 (study baseline) for exposure assessment.

 

%% file: sec2.tex
\section{Methods}
\label{sec:meth}
 
\subsection{Functional linear Cox regression model}
We let $\tilde{T}_i$ be time to event of interest, $C_i$ the censoring time for individual $i=1, \ldots, n$, and denote $\tau$ as the end of follow-up time. We define $X_i(s)$ as the average of environmental exposure on a concentric ring at radius $s$, centered on the residence for individual $i$. We let $\mathcal{S}$ be a closed and bounded interval of $\mathbb{R}$ and $\{X_i(s): s\in \mathcal{S}\} $ be a random function expressing the exposure.  We take $\mathcal{S}$ to be $[0, 1]$ without loss of generality. 
We observe the environmental exposures at the pre-specified radii of concentric rings $\{s_{r} \in \mathcal{S}, 1\leq r \leq  \mathcal{R} \}$, where $\mathcal{R}$ is the number of radii of observed concentric rings.  
Let $Z_i = (Z_{i1}, \ldots, Z_{ip})^{T}$ be a $p$-vector of scalar covariates that we adjust for, and define $W_i = ( X_i, Z_i)^{T}$. Under the non-informative right-censoring setting, we observe $n$ i.i.d. samples $(T_i, \Delta_i, Z_i, \{X_i(s_r), 1 \leq r \leq \mathcal{R} \})$ for $i=1, \ldots, n$, where $T_i = \min (\tilde{T}_i, C_i)$, and $\Delta_i=I(T_i \leq C_i)$ is an indicator for occurrence of the event. The functional linear Cox regression model specifies the hazard function as
\begin{equation}\label{eq1}
\lambda(t) = \lambda_{0}(t)\exp\left(\int_0^{1} X_i(s)\beta(s)ds + \theta^{T}Z_i\right),
\end{equation}
where $\lambda(t) = d\Lambda(t)$, $\lambda_{0}(t) = d\Lambda_0(t)$ is the baseline hazard function, $\beta(s)$ represents the effect of average environmental exposures $X(s)$ measured on the concentric rings with radius $s$ on the survival outcome, and we denote $\theta = (\theta_1, \ldots, \theta_p)^{T}$ for a vector of covariate effects. 

Our approach utilizes the B-spline basis expansion of $\beta(s)$ \citep{deboor2001spline, james2009functional, lin2017locally}. We let $\mathcal{S}_{dM_n}$ denote the linear space of spline functions on $\mathcal{S}$ spanned by the B-spline basis functions $\{B_j(s): j=1, \ldots, M_n + d + 1\}$ with degree $d$ and   $M_n$ equally spaced inner knots $0=\kappa_0<\kappa_1<\cdots <\kappa_{M_n}<\kappa_{M_n+1} =1$, and let $B(s) = (B_{1}(s), \ldots, B_{M_n+d+1}(s))^{T}$.  Over each of subintervals $[s_{m}, s_{m+1}]$, each B-spline basis function $B_{m}(s)$ is a piecewise polynomial of degree $d$ and nonzero over no more than $d+1$ adjacent subintervals. 
Under suitable smoothness assumptions, we can approximate $\beta(s)$ as a linear combination of B-spline basis functions 
$\beta(s)\approx \beta^*(s) = \sum_{k=1}^{L_n} b_k B_k(s) = B^{T}(s)b$,
where $L_n = M_n + d+ 1$, $\beta^*(s) \in \mathcal{S}_{dM_n}$, and $b = (b_1, \ldots, b_{L_n})^{T}$ is a vector of coefficients. 

\subsection{Smoothness and sparsity of $\beta(s)$}

The estimators of coefficient functions $\beta(s)$ can be wiggly when $M_n$ becomes large. To enforce smoothness on $\beta(s)$, we impose the smoothness penalty  $b^{T}Jb$ where $J_{i,j} = \int_0^1 \frac{\partial^2{B}_{i}(s)}{\partial s^2} \frac{\partial^2{B}_{j}(s)}{\partial s^2} d s$  \citep{cardot2003spline, yuan2010reproducing}. 

In order to identify the buffer distance, we need a sparse representation of $\beta(s)$. By obtaining the sparsity of $\beta(s)$, it is possible to identify non-null regions $\mathcal{S}_{1}=\{s: \beta(s)\neq 0, s \in \mathcal{S} \}$ and null regions $
\mathcal{S}_{0}=\{s: \beta(s)= 0, s \in \mathcal{S} \}$. 
By the B-spline approximation, the sparsity of $\beta(s)$ can be characterized by the sparsity of the coefficient vector $b$. Specifically, if $\beta(s)=0$ for $s\in [\kappa_{j-1}, \kappa_j]$, then $b_j = \cdots = b_{j+d}=0$ since the B-spline basis functions have the local support property \citep{wang2015functional}. This overlapping group structure of coefficients of $b$ requires the selection of groups of variables where the groups can be overlapped. In this regard, we impose the group bridge penalty \citep{huang2009group, huang2014group} to select at both group-level and individual-level. 
Therefore, by estimating $b$, we can estimate $\beta(s)$ and identify the buffering distance. 

We estimate $\alpha = (b^{T}, \theta^{T})^{T}$ by maximizing the penalized log partial likelihood
\begin{equation}\label{eq2}
\mathcal{L}_n(\alpha)  =n^{-1} l_n(\alpha) - \lambda_{1} \sum_{j=1}^{M_n+1} \|b_{A_j} \|_1^{\gamma} -\lambda_{2}b^T Jb ,  
\end{equation}
where $l_n(\alpha)$, the log partial likelihood, is given by
\begin{align}\label{eq3}
l_n(\alpha) = & \sum_{i=1}^n  \Delta_i\bigg\{ \sum_{k=1}^{L_n} b_k  \int_0^{1} X_i(s)B_k(s)ds  + \theta^{T} Z_i  \notag \\&- \log \sum_{T_j \geq T_i} \exp\left( \sum_{k=1}^{L_n} b_k  \int_0^{1} X_j(s)B_k(s)ds  + \theta^{T}Z_j\right)\bigg\},
\end{align}
 $A_j= \{j, j+1, \ldots, j + d\}$, $b_{A_j} = (b_k, k \in A_j)^T$, and $\| b_{A_j} \|_1 = |b_j| + \ldots + |b_{j+d}|$ for $0<\gamma<1$  \citep{wang2015functional}. We note that $\lambda_1$ is a regularization parameter that controls the local sparsity of $\beta(s)$, and $\lambda_{2}$ enforces smoothness on $\beta(s)$. We denote the maximizers as $\hat{\theta}$ and $\hat{b}$, then the spline estimator of $\beta$ is $\hat{\beta}(s) = B(s)^{T}\hat{b}$.

\subsection{Computational algorithm}\label{Sec3.4}
Since direct maximization of \eqref{eq2} is difficult due to the non-convexity of the group bridge penalty for $0<\gamma<1$, we follow the approach of \cite{huang2009group} to maximize \eqref{eq2}. For $0<\gamma<1$, we let 
\begin{equation*}
\mathcal{L}_n(\alpha) = n^{-1}l_n(\alpha) - \sum_{j=1}^{M_n+1}\mu_j^{1-1/\gamma} \|b_{A_j}\|_1 - \zeta \sum_{j=1}^{M_n+1} \mu_j -\lambda_{2}b^T Jb,
\end{equation*}
where $\mu_j = \{(1-\gamma)/(\gamma \zeta)\}^{\gamma} \| b_{A_j} \|_1^{\gamma}$ with $\zeta = \lambda_1^{1/(1-\gamma)}\gamma^{\gamma/(1-\gamma)}(1-\gamma).$ Define $\tilde{l}_n(\alpha) = - l_n(\alpha)$
and denote the gradient vector $\bigtriangledown \tilde{l}_n(\alpha) = \partial \tilde{l}_n(\alpha)/\partial \alpha$ and the Hessian matrix $\bigtriangledown^2 \tilde{l}_n(\alpha) = \partial^2 \tilde{l}_n(\alpha)/\partial \alpha\partial \alpha^T$. Consider the Cholesky decomposition of $\bigtriangledown^2 \tilde{l}_n(\alpha^\circ), \bigtriangledown^2 \tilde{l}_n(\alpha^\circ) = V^{T}V$ ($V$ is an upper triangle matrix), and set the pseudo response vector $Y = (V^{T})^{-1} \{\bigtriangledown^2 \tilde{l}_n(\alpha^\circ)\alpha^\circ - \bigtriangledown \tilde{l}_n(\alpha^\circ)\}$. Following the arguments in \cite{zhang2007adaptive} and \cite{huang2014group}, the partial log-likelihood can be approximated by a quadratic form using a Taylor expansion around $\alpha^\circ$, i.e.,  $\tilde{l}_n(\alpha) \approx 1/2 (Y-V\alpha)^{T}(Y-V\alpha)$ (see Section 1 of the Supplemental Materials for the derivation). Therefore, we propose  to minimize 
\begin{equation}\label{eq4}
\frac{1}{2n} (Y-V\alpha)^{T}(Y-V\alpha) + \sum_{j=1}^{M_n+1}\mu_j^{1-1/\gamma} \|b_{A_j}\|_1 + \zeta\sum_{j=1}^{M_n+1} \mu_j +  \lambda_{2}b^T Jb.
\end{equation}
 Let $V = (v_1 \cdots v_{L_n}, v_{L_n+1} \cdots v_{L_n+p})^T$, and define an augmented data set  $(\overline{Y}, \overline{V})$ \citep{zou2005regularization} as 
\begin{equation*}
\overline{Y} = \begin{pmatrix}
Y\\
\mathbf{0}
\end{pmatrix},
\quad \quad 
\overline{V} = \begin{pmatrix}
v_1\cdots v_{L_n} & v_{L_n+1}\cdots v_{L_n+p}  \\
\sqrt{2n\lambda_2}D & \mathbf{0}^T
\end{pmatrix},
\end{equation*}
where $J = D^{T}D.$ 
It follows that, \eqref{eq4} can be written as
\begin{equation}
\mathcal{Q}_n(\alpha) =  \frac{1}{2n} (\overline{Y} -\overline{V} \alpha)^{T} (\overline{Y} - \overline{V} \alpha) + \sum_{j=1}^{M_n+1}\mu_j^{1-1/\gamma} \|b_{A_j}\|_1 + \zeta\sum_{j=1}^{M_n+1} \mu_j.
\end{equation}
We propose the following algorithm:
\begin{itemize}[
align=left,
  leftmargin=2em,
  itemindent=2pt,
  labelsep=0pt,
  labelwidth=2em,
  itemsep=2mm
]
\raggedright
\item[Step 1.  ] For a given choice of $\lambda_2$, obtain an initial value of $\alpha$, denoted by $\alpha^{(0)}$, by solving  $ \argmin\limits_{\alpha} \frac{1}{2n} (Y - V\alpha )^{T} (Y - V\alpha) +    \lambda_2b^{T} Jb $ iteratively until convergence where an initial value is a $\mathbf{0}$ vector \citep{gellar2015cox}.  The solution at each iteration is given by 
$$\hat{\alpha} = (V^T V + 2n \lambda_2 J^*)^{-1}V^T Y$$
where $J^*$ is a block diagonal matrix with $J$ and the zero matrix of size $p \times p$ on the diagonal.  
 \item[Step 2.  ]   At the $m$th iteration, $m=1,2,\ldots$, compute 
$$\mu_j^{(m)} =\left( \frac{1-\gamma}{\gamma \zeta}\right)^{\gamma} \| b_{A_j}^{(m-1)} \|_1^{\gamma}, \quad j=1, \ldots, M_n+1,$$
and 
$$\xi_k^{(m)} = \sum_{j:k\in A_j} (\mu_j^{(m)})^{1-1/\gamma}, \quad k=1, \ldots, L_n.$$
 \item[Step 3.  ] Compute $\overline{Y}$ and $\overline{V}$ based on $\alpha^{(m-1)}$.
 \item[Step 4. ] Obtain the new estimate by coordinate descent algorithm as
$$\alpha^{(m)} = \argmin_{\alpha} \left\{  \frac{1}{2n} (\overline{Y} -\overline{V} \alpha)^{T} (\overline{Y} - \overline{V} \alpha)  + \sum_{k=1}^{L_n} \xi_k^{(m)} |b_k|\right\}.$$
  \item[Step 5.  ] Stop with convergence. 
\end{itemize}
The algorithm always converges to a local minimizer of $\mathcal{Q}(\alpha)$ given fixed tuning parameters depending on the initial value of $\alpha^{(0)}$, since the group penalty is not convex. 
To avoid the effect of this issue, we suggest to use the multiple initial values and select the one which can achieve the smallest value of $\mathcal{Q}(\alpha)$.

%% file: sec3.tex
\subsection{Assumptions and asymptotic results} 

In this section, we discuss the asymptotic properties of the estimators defined in the previous section.
 We first introduce notations. Let $(\beta_0, \theta_0)^{T}$ denote the true parameter value. 
 We let $\|\cdot\|_2$ denote the Euclidean norm and  $\|\cdot\|_{\infty}$ denote the supremum norm. In the Hilbert space $H = L^2(\mathcal{S})$, containing all square integrable functions on $\mathcal{S}$, we define inner product $\inner{x}{y} = \int_{\mathcal{S}} x(s)y(s)ds$, $\forall x, y \in H$ and norm $\|x \| = \inner{x}{x}^{1/2}$.
For two sequences of positive numbers $a_n$ and $b_n$, $a_n\lesssim b_n$ indicates $a_n/b_n$ is uniformly bounded and $a_n \asymp b_n$ if $a_n\lesssim b_n$ and $b_n\lesssim a_n$. 
We define the covariance operator $\Gamma$ of the random process $X$ as $\Gamma x(t) = \int_0^{1} EX(t)X(s)x(s)ds$, and $\Gamma_{\Delta} x(t) = \int_0^{1} E\Delta X(t)\Delta X(s)x(s)ds =  \int_0^{1} E\Delta X(t)X(s)x(s)ds$. We denote $\Gamma_n$ as the empirical version of $\Gamma$ of X; that is, $\Gamma_n x(t) = n^{-1} \sum_{i=1}^n \int_0^1 X_i(t)X_i(s)x(s)ds.$  Let $H$ be the $n \times M_n$ matrix with element $h_{i,j} = \inner{\Gamma_n B_i} {B_j}$. 
Let $\rho_{min}(A)$ and $\rho_{max}(A)$ denote the minimum and maximum eigenvalues of a 
symmetric matrix $A$, respectively.  For a $s\times s$ matrix $A$, $\|A\|_{\infty} = \max\{\sum_{k=1}^s |(A)_{jk}|: 1\leq j \leq s\}$ and  $\|A\|_2 = \{\rho_{\max} (A^{T}A)\}^{1/2}$. We define the seminorm $\| f \|_{\Gamma} = \left\{ \int_0^{1} (\Gamma f)(s)f(s)ds\right\}^{1/2}$.  
Let $\mathcal{H}(r)$ be the collection of functions defined on $[0,1]$ whose $v$th derivative exists and satisfies the H\"older condition of order $\delta$ with $r \equiv v + \delta > 2$, $|f^{(v)}(s_1) - f^{(v)}(s_2)| \leq D|s_1 - s_2|^{\delta}$, for $0 \leq s_1, s_2 \leq 1$ and some constant $D>0$.

We make the following assumptions: 
\begin{enumerate}
\item[(A1)] $\beta \in \mathcal{H}(r)$.
\item[(A2)] Assume that the exposure function and the additional covariates have been centered so that $E(\Delta Z) = 0$ and $E(\Delta X(s)) = 0, s\in [0, 1]$.
\item[(A3)] Assume $\tilde{T} \perp C | W$.
\item[(A4)] The observed event time $T_i$, $1 \leq i \leq n$ is in a finite interval, say $[0, \tau]$, and the baseline cumulative hazard function $\int_0^{\tau} \lambda_0(s)ds < \infty$, and there exists a small positive constant $\epsilon$ such that: (i) $P(\Delta = 1 | W) > \epsilon$, and (ii) $P(C>\tau | W) > \epsilon$ almost surely with respect to the probability measure of $W$.
\item[(A5)] The covariate $Z$ takes values in a bounded subset of $R^{p}$, and the $L_2$-norm $\|X\|_2$ is bounded almost surely. 
\end{enumerate}

To have a sparse estimator of $\beta(s)$, we partition the set of number of knots $\{1, \ldots, M_n+1\}$ into three sets \citep{wang2015functional}
\begin{align*}
\mathcal{A}_1 &= \{j : \beta(s) = 0, s \in [\kappa_{j-1}, \kappa_j] \},\\
\mathcal{A}_2 &= \{j : 0 < \max_{s\in [\kappa_{j-1}, \kappa_j]} |\beta(s)| \leq D^*M_n^{-r}\text{ for some } D^*>D\},\\
\mathcal{A}_3 &= \{1, \ldots, M_n+1\} - \mathcal{A}_1\cup  \mathcal{A}_2.
\end{align*}
If (A1) holds, there is a vector of $\beta(s)^*=B^{T}(s)b^*$ where $\left\|\beta_0(s)-\beta^*(s)\right\|_{\infty} \leq D_0 M_n^{-r}$ for some constant $D_0$. As suggested in \cite{wang2015functional}, we introduce a sparse modification of $\beta^*(s)\in  \mathcal{S}_{dM_n}$ denoted by $\beta_0^*(s) = B^{T}(s)b_{0n}$, where $b_{0n} = (b_{0,1}, \ldots, b_{0,L_n})^{T}$ is a sparse modification of $b^*$ with $b_{0,k} = b_{k}^*I(k \not\in \mathcal{B}_1)$, and $\mathcal{B}_1 = \cup_{j\in \mathcal{A}_1 \cup \mathcal{A}_2} A_j$. Note that under (A1), $\| \beta_0 - B^{T}b_{0n} \|_{\infty} = O(M_n^{-r})$ by Lemma 1 in  \cite{wang2015functional}.

\begin{enumerate}
\item[(A6)] $M_n = o(n^{1/4})$, $\lambda_{1}\nu_n = O(n^{-1/2}M_n^{1/2})$ with $\nu_n =\left(\sum_{j \in \mathcal{A}_3} \|b_{0, A_j} \|_1^{2\gamma-2}\right)^{1/2}$ and $\lambda_{2} =o(n^{-1/2}M_n^{1/2})$. 
\end{enumerate}

The assumptions (A7) -- (A10) are in the Supplementary Materials. 
The assumptions (A1) -- (A5), and (A7) -- (A8) are the same as those in  \cite{qu2016optimal} and the assumptions (A6), (A9) and (A10) are similar to \cite{huang2014group}.

 \begin{theorem}[Convergence rate] \label{thm1} Define $D_n = M_n^{1/2} n^{-1/2} + M_n^{-r}$. Under conditions (A1) -- (A10), there exists a local maximizer $(\hat{b}, \hat{\theta})$ such that 
 $$\| \hat{\theta} - \theta_0 \|_2= O_p(D_n),$$
 $$\| \hat{\beta} - \beta_0 \|= O_p(D_n),$$
 and
 $$ \| \hat{\beta} - \beta_0 \|_{\Gamma}= O_p(D_n).$$
 \end{theorem}
By Theorem 1, the optimal choice of $M_n \asymp O(n^{1/(2r+1)})$.  Then, $\| \hat{\theta} - \theta_0 \|_2 = O_p(n^{-r/(2r+1)})$ and $\| \hat{\beta} - \beta_0 \|_{\Gamma} = O_p(n^{-r/(2r+1)})$. 
The details of the proof are provided in the Supplementary Materials, which show that there exists a local maximum in the ball $\{\alpha_0 + \delta_n u: \| u \|_2 = c_0\}$ with probability at least $1-\epsilon$, where $\alpha_0 = (b_{0n}, \theta_0)$. 

 \begin{theorem}[Sparsistency] Under conditions (A1) -- (A10), we have $(\hat{b}_{A_j} : j \in \mathcal{A}_1 \cup \mathcal{A}_2 ) = \bm{0}$ with probability tending to 1. 
 \end{theorem}
 This theorem guarantees that the non-null regions are identified by a set of $\mathcal{A}_1 \cup \mathcal{A}_2$, in which $\hat{\beta}(s)=0$ with large probabilities when $s \in \cup_{j: \mathcal{A}_1 \cup \mathcal{A}_2} [\kappa_{j-1}, \kappa_j]$.  The details of the proof are provided in the Supplementary Materials.

\subsection{Choice of tuning parameters}\label{Sec3.5}

The choice of tuning parameters determines the performance of the proposed methods. We choose the number of interior knots $M_n$ to be relatively large, following \cite{marx1999generalized, lin2017locally}. Alternatively, we could use K-fold cross-validation to choose $M_n$ \citep{huang2004polynomial, wang2015functional}. As suggested in \cite{huang2010variable}, we set $\gamma=0.5.$ Using the techniques similar to those in \cite{guan2020estimating}, we compute the effective number of parameters by the Newton-Raphson method, given by
\begin{equation*}
df(\lambda_1, \lambda_2) = \text{tr}[ \{H_{0} + n\lambda_2J_{0} \}^{-1} H_{0}],
\end{equation*}
where $H_{0}$ denote the submatrix of  $\bigtriangledown^2 l(\hat{\alpha})$ with rows and columns corresponding to the nonzero $\hat{b}$, and $J_{0}$ denote the submatrix of $J$ with rows and columns corresponding to the nonzero $\hat{b}$.
We select tuning parameters $\lambda_1$ and $\lambda_2$ using a BIC-type criterion,
\begin{equation}\label{eq6}
\text{BIC}(\lambda_1, \lambda_2) = -2 \log l_n(\hat{\alpha})  + \log(n)df(\lambda_1, \lambda_2).
\end{equation}
The BIC-type criterion is better suited to identify the true model structure because it penalizes a model more for its complexity compared to AIC \citep{bishop2006pattern}.

\subsection{Variance estimation using two-stage approach}\label{Sec3.6}

In this section, we introduce a  two-stage approach to obtain the interval estimates for $\beta(t)$ given the non-null regions. In the first stage, we define the non-null region indicators as $I_0 = \bigcup_{j=1}^{M_n+1}\{j: \hat{\beta}(s)\neq 0 \ \text{for}\ \forall s\in [\kappa_{j-1}, \kappa_j]\}$, and then non-null regions are $\mathcal{S}_s = \bigcup_{j \in I_0} [\kappa_{j-1}, \kappa_j]$, where $\hat{\beta}(s)$ on $\mathcal{S}$ are estimated based on the algorithm in Section \ref{Sec3.4}. We define $A_s =  \bigcup_{j \in J_0} A_j$.  
In the second stage, given the identified non-null regions $\mathcal{S}_s$,  we re-estimate $\beta(s)$ in $s \in \mathcal{S}_s$, denoted by $\hat{\beta}^*(s)$, based on the B-spline basis functions $\{ B_j(s): j\in A_s \},$ by maximizing the penalized partial likelihood only with the smoothness penalty, written as
\begin{equation}\label{eq7}
\mathcal{L}^\ast_n(\tilde{\alpha})  =n^{-1} l_n^*(\tilde{\alpha})  -\tilde{\lambda}_{2}\tilde{b}^T J_{0}\tilde{b},  
\end{equation}
where the log partial likelihood is given by
\begin{align*}
l_n^\ast(\tilde{\alpha}) = & \sum_{i=1}^n  \Delta_i\bigg\{ \sum\limits_{k\in A_s} b_k  \int_0^{1} X_i(s)B_k(s)ds  + \theta^{T} Z_i  \notag \\&- \log \sum_{T_j \geq T_i} \exp\left( \sum\limits_{k \in A_s} b_k  \int_0^{1} X_j(s)B_k(s)ds  + \theta^{T}Z_j\right)\bigg\},
\end{align*}
$\tilde{b} = (b_k, k\in A_s)^T$, and $\tilde{\alpha} =(\tilde{b}^T, \theta^T)^T$. Note that coefficients corresponding to the B-spline basis functions are included only in the non-null regions.  
Here, the smoothness penalty tuning parameter $\tilde{\lambda}_2$ is chosen by a BIC-type criterion of the form \eqref{eq6}. 

Based on the Condition (C2) in \citet{hao2021semiparametric}, we simultaneously diagonalize $H =  -\partial^2 l^*_n(\tilde{\alpha})/\partial \tilde{\alpha}\partial \tilde{\alpha}^T$ and $$\mathcal{P} =\begin{bmatrix} J_0 & 0_{q\times p} \\  0_{p \times q} & 0_{p \times p} & \end{bmatrix}$$ by producing a matrix $R$ and a diagonal matrix $\mathcal{D}$ such that 
$R^{T} H R = I$ and $R^T \mathcal{P} R = \mathcal{D}$. 
Now, we write $R$ and $\mathcal{D}$ as 
$$R = \begin{bmatrix}
    R_{11} & R_{12} \\
    R_{21} & R_{22}
\end{bmatrix}, \quad \mathcal{D} = \begin{bmatrix}
    \Pi & 0_{q \times p} \\ 0_{p\times q} & 0_{p\times p} 
\end{bmatrix},$$
where $\Pi$ is a diagonal matrix with diagonal elements $\pi_1, \ldots, \pi_{q}$ with an increasing order.
Then, we define
$$\psi_l(s) = \sum_{u=1}^q R_{12,ul}B_u(t),$$
$$\phi_{\nu}(s) = \sum_{u=1}^q R_{11,u\nu}B_u(t).$$
By applying the Theorem 3.3 in \cite{hao2021semiparametric}, 
\begin{equation}\label{eq8}
\widehat{\operatorname{Var}}(\hat{\beta}^*(s))=\frac{1}{n}\left[\sum_{l=1}^p \psi_l^2(s)+\sum_{\nu=1}^q \frac{\varphi_\nu^2(s)}{\left(1+\tilde{\lambda}_2 \pi_\nu\right)^2}\right],
\end{equation}
where $q=|A_s|$. 

It is often of interest to estimate cumulative effects $\int_{s\in \mathcal{S}_s} \hat{\beta}^*(s) ds$, which summarizes the overall effect of exposure within a circular distance buffer. We can calculate the variance of the cumulative effect as
$$\widehat{\operatorname{Var}}\left(\int_{s\in \mathcal{S}_s}  \hat{\beta}^*(s) ds \right) =\frac{1}{n}\left[\sum_{l=1}^p |\psi_l^0|^2+\sum_{\nu=1}^q \frac{ |\varphi_\nu^0|^2}{\left(1+\tilde{\lambda}_2 \pi_\nu\right)^2}\right],$$
where $\psi_l^0 = \int_{s\in \mathcal{S}_s} \psi_l(s)ds$, and $\phi_\nu^0 =  \int_{s\in \mathcal{S}_s} \phi_\nu(s)ds$.

Note that this variance formula is only approximate because we do not take into account the variance associated with the selection of the nonzero $\beta(s)$ in the first stage. In our simulation studies, we observed that this approximation performs well.

%% file: sec4.tex
\section{Simulation Studies}\label{sec:simu}

We conducted simulation studies to evaluate the finite sample performance of the proposed method (denoted by \texttt{Spline-Gbridge}), compared with other methods including the \texttt{Spline} method \citep{gellar2015cox},  the \texttt{Lasso} method, the \texttt{Gbridge} method \citep{huang2009group} and the \texttt{Spline-Lasso} method \citep{guo2016spline}. The \texttt{Spline} method has the smoothness penalty and no sparseness penalty in \eqref{eq2} by setting $\lambda_1=0$.  The \texttt{Lasso} method has the lasso penalty where the group bridge penalty is replaced with $\lambda_1\sum_{j=1}^{L_n} |b_j|$ and no smoothness penalty in \eqref{eq2} by setting $\lambda_2=0$. The \texttt{Gbridge} method has the group bridge penalty and no smoothness penalty in \eqref{eq2} by setting $\lambda_2=0$.  The \texttt{Spline-lasso} method has the smoothness penalty and the lasso penalty in \eqref{eq2}.

We generated failure time $\tilde{T}_i$  from an exponential distribution with $\lambda_0(s) = 1$ and  a hazard function
\begin{equation*}
\lambda(t) = \lambda_{0}(t)\exp\left(\int_0^{1} X_i(s)\beta(s)ds + \theta_1Z_{i1} + \theta_2 Z_{i2} \right),
\end{equation*}
where the covariate function $X_i(s)$ was generated by $X_i(s) = \sum_{j=1}^{52} c_{ij}B_j(s)$,  where $B_j(s)$ are cubic B-spline basis functions with 48 equally spaced inner knots, and $c_{ij}$ was generated from the standard normal distribution. Two scalar covariates were generated by $Z_1 \sim N(1,0.5^2)$ and $Z_2 \sim N(0, 1^2)$, respectively. We set $\theta_1 = \log(0.8)$, and $\theta_2 = \log(1.2).$ 
We consider three scenarios for $\beta(s)$ with $s\in [0,1]$: 
\begin{itemize}
\item Scenario I: $\beta_1(s) = 0$,
\item Scenario II: $\beta_2(s) = 2\sin(2\pi s) I_{0\leq s< 0.5}$,
\item Scenario III:  $\beta_3(s) = -2\sin(\pi(s-0.5))I_{0\leq s < 0.5}$,
\end{itemize}
where $I_{(\cdot)}$ denotes the indicator function. For the scenarios II and III, $\beta(s)=0$ on $[0.5, 1]$.  
The censoring time was generated from an exponential distribution with a censoring rate of $10\%$. 

In the fitted models, we set $M_n=26$ and $d=3$. We considered sample sizes of $n=500$ and $1000$, and we independently generated $1000$ datasets for each case. The performance of estimators $\hat{\beta}(s)$ was evaluated by the integrated mean squared errors according to IMSE = $\int_0^1 \left(\hat{\beta}(s) - \beta(s)\right)^2 ds/\int_0^1 \beta(s)^2 ds$ for Scenario II and III, and we consider the IMSE $\int_0^1 \left(\hat{\beta}(s) - \beta(s)\right)^2 ds$ for  Scenario I. We also examined the performance of estimators $\hat{\theta}_1$ and $\hat{\theta}_2$ with percent bias and empirical standard errors (in parentheses). 
To assess the performance of $\hat{\beta}(s)$ on identifying the non-null regions, equivalently the supremum of non-null regions in our scenarios, we obtained the average of the supremum of non-null regions.  The tuning parameters $\lambda_1$ and $\lambda_2$ were selected based on BIC. 

\begin{table}[h!]
\caption{Simulation results for three scenarios. The mean and standard deviation (in parentheses) of IMSE for $\hat{\beta}(s)$ and a percent bias and empirical standard error for $\hat{\theta}_1$ and  $\hat{\theta}_2$. Tuning parameters are selected based on BIC. We set a sample size of $n=1000$, and the results are based on 1000 simulated data sets. The results for $n=500$ are shown in the Supplementary Materials.}
\label{Table1}
\vspace{3mm}
\resizebox{\columnwidth}{!}{%
\begin{tabular}{ccccccc}
\hline
Scenario & Parameter&  Spline & Lasso & Gbridge &  Spline-Lasso &  Spline-Gbridge\\
\hline 
\multirow{3}{*}{I} & $\beta_1(s)$&  0.154 (0.155) & 0.010 (0.056)  & 0.084 (1.199) & 0.010 (0.062) &  0.046 (0.131) \\
& $\theta_1$   & -0.145 (0.076) & -0.287 (0.076)  & -0.325 (0.076) & -0.301 (0.076) &  -0.293 (0.076) \\
&$\theta_2$ & -0.539 (0.038) & -0.704 (0.038)  & -0.675 (0.038) & -0.718 (0.038) &  -0.642 (0.038) \\


\multirow{3}{*}{II}&$\beta_2(s)$ &  1.160 (0.483) & 0.907 (0.162) & 0.947 (1.265) & 0.527 (0.348) & 0.483 (0.344) \\
& $\theta_1$  & -0.188 (0.076) & -0.840 (0.077)  & -0.662 (0.078) & -0.385 (0.078) &  -0.221 (0.077) \\
& $\theta_2$   & -0.886 (0.038) &-1.585 (0.038)  & -1.343 (0.038) & -1.074 (0.038) &  -0.909 (0.038) \\

 
\multirow{3}{*}{III}&$\beta_3(s)$&  0.237 (0.155) & 0.931 (0.134)  & 0.971 (0.512) & 0.458 (0.378) & 0.416 (0.387)  \\
&$\theta_1$ & -0.027 (0.077) & -0.840 (0.078)  & -0.714 (0.078) & -0.366 (0.077) &  -0.265 (0.077) \\
&$\theta_2$ & -0.757 (0.038) & -1.585 (0.038)  & -1.426 (0.038) & -1.091 (0.038) &  -1.011 (0.038) \\
\hline
\end{tabular}}
\end{table}

Table~\ref{Table1} reports the mean and the standard deviation (in parentheses) of IMSE when the sample size is $n=1000$.  
 The estimators based on the \texttt{Spline-Gbridge} model outperform over those based on the \texttt{Gbridge} model. 
 For Scenario I where no effect exists, the \texttt{Lasso} and \texttt{Spline-Lasso} models give smaller IMSE than the other models. However, for Scenario II and III, the \texttt{Spline-Gbridge} estimators have the smallest IMSE than the other methods. The simulation results for $n=500$ are in the Supplementary Material. Overall, our proposed model \texttt{Spline-Gbridge} gives better performance in terms of IMSE.
  
 Table~\ref{Table2} reports the mean and standard deviation (in parentheses) of the supremum of the non-null region, i.e., the buffer distance. In the true model, the buffer distance is 0 for Scenario I and 0.5 for  Scenario II and III. The \texttt{Spline} model performs well in terms of IMSE but it cannot identify the buffer distance.  The performance of identifying the buffer distance based on the \texttt{Lasso}  and \texttt{Spline-Lasso} for  Scenario II and III is poor. The simulation results for $n=500$ are in the Supplementary Materials. The \texttt{Spline-Gbridge} estimators outperform the other methods in terms of estimation accuracy as well as identifying the buffer distance by exploiting the local support property of B-spline basis functions \citep{wang2015functional}.   
 
\begin{table}[h!]
\centering
\caption{Simulation results for three scenarios. The average of the supremum of the non-null regions. 
Tuning parameters are selected based on BIC. We set a sample size of $n=1000$, and the results are based on 1000 simulated data sets. The results for $n=500$ are shown in the Section 3 in the Supplementary Materials.}
\label{Table2}
\vspace{3mm}
\begin{tabular}{ccccc}
\hline
Scenario  & Lasso & Gbridge & Spline-Lasso & Spline-Gbridge\\
\hline


I &  0.058 (0.199) & 0.040 (0.166) & 0.041 (0.177) & 0.109 (0.260)  \\

II& 0.136 (0.208) & 0.208 (0.223) & 0.683 (0.429) & 0.563 (0.298)   \\

III & 0.091 (0.172) & 0.124 (0.192) & 0.683 (0.445) & 0.597 (0.342)  \\
\hline
\end{tabular}
\end{table}

We conducted additional simulation studies to assess the accuracy of the estimate introduced in Section~\ref{Sec3.6} of the standard error of $\hat{\beta}^*(s)$. Therefore, we conducted the simulation studies with Scenarios II and III given the true non-zero areas $[0, 0.5]$ as follows. 
Given the true non-zero areas $[0, 0.5]$, we estimate $\beta(s)$ in $[0, 0.5]$, denoted by $\hat{\beta}^*(s)$, by maximizing \eqref{eq7} and estimate the standard error of $\hat{\beta}^*(s)$ using the formula \eqref{eq8}. The smoothness penalty tuning parameter was chosen by the BIC-type criterion. 
\renewcommand{\thefigure}{2}
\begin{figure}[h!]
\centering
\includegraphics[scale=0.7]{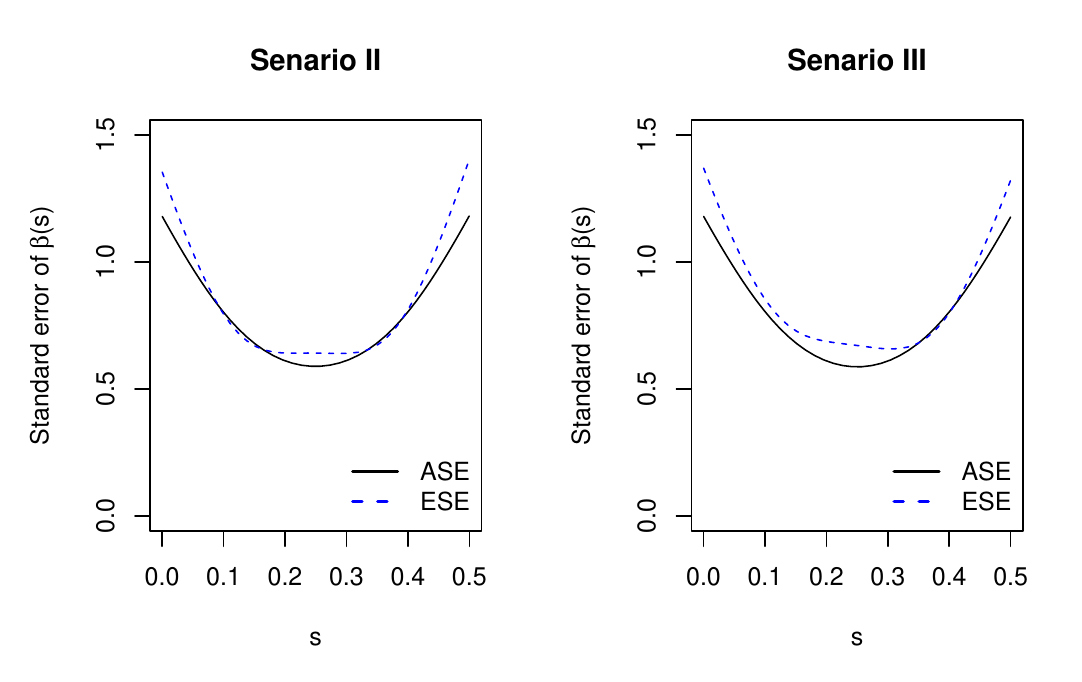}
\caption{Simulation results for two-stage variance estimation for Scenarios II and III on the true non-zero areas $[0, 0.5]$. The black solid lines represent the average of the estimated standard error (ASE), and the blue dotted lines represent the empirical standard error (ESE) for Scenario II (the left penal) and Scenario III (the right penal).}
\label{Figure2}
\end{figure}

\renewcommand{\thefigure}{3}
\begin{figure}[h!]
\centering
\includegraphics[scale=0.7]{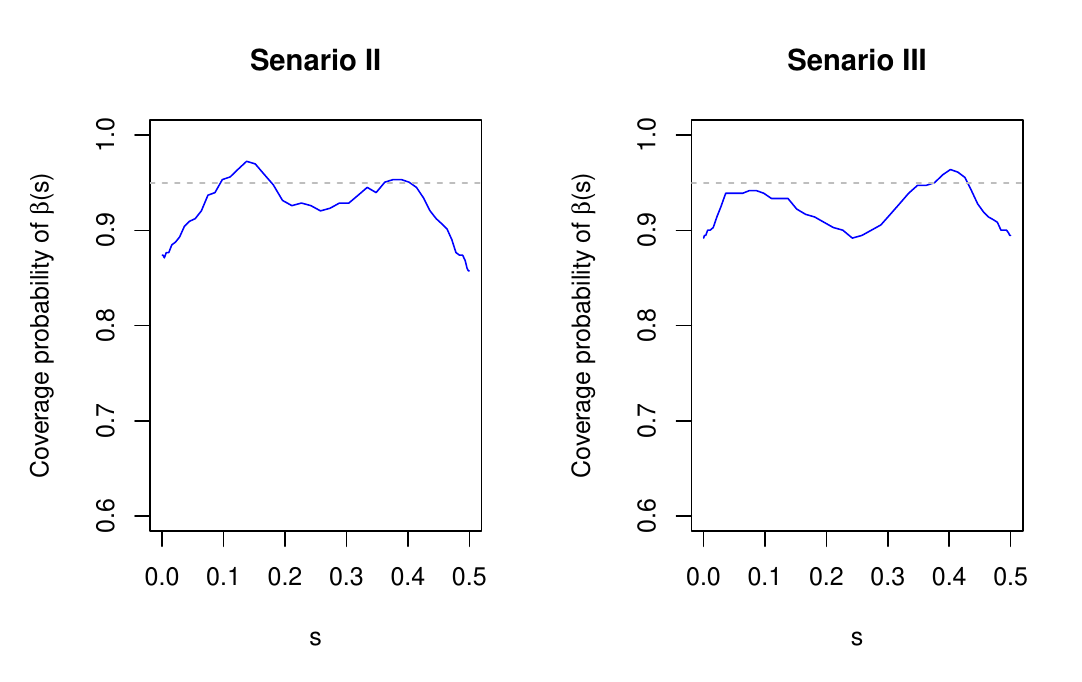}
\caption{Simulation results for two-stage variance estimation for scenarios II and III on the true non-zero areas $[0, 0.5]$. The left penal shows the coverage probability for $\beta(s)$ for Scenario II and the right penal shows the the coverage probability for $\beta(s)$ for Scenario III.}
\label{Figure3}
\end{figure}

Figure~\ref{Figure2} displays the average of the estimated standard error (ASE) of $\beta(s)$ and the empirical standard error (ESE) of $\beta(s)$.  The average of the estimated standard errors and the empirical standard erros are in good agreement in both Scenarios. Figure~\ref{Figure3} illustrates the coverage probability of $\beta(s)$. Scenario II shows a slight undercoverage at the end and the center of the interval, however, overall, it shows good performance. For Scenario III, we see a generally good performance over the interval. 

%% file: sec8.tex
\section{Case Study}\label{sec:data}

The aim of the case study is to investigate the radial buffer distance of NDVI after which there is no effect on depression incidence and to estimate the effects of green space on depression incidence in NHS within the selected buffer distance. 
The objective measurements of green space consist of the NDVI data assessed within circular distance buffers at nine radii ($s_1$=90m, $s_2$=150m, $s_3$=270m, $s_4$=510m, $s_5$=750m, $s_6$=990m, $s_7$=1230m, $s_8$=1500m, and $s_9$=2100m). We obtained the average NDVI values in the distance buffers between radii $s_{j-1}$ and $s_{j}$, denoted by $X_i(s_{j})$, for $j=1, \ldots, 9$ and we let $s_0=0$. 
Figure~\ref{Figure4} shows the pattern of the average NDVI values in concentric rings based on circular distance buffers at radii of 0m, 90m, 150m, 270m, 510m, 750m, 990m, 1230m, 1500m, and 2100m for a total of 35,283 eligible participants in NHS. The range of the average NDVI was $(0.48, 0.49)$ and the maximum NDVI of 0.49 was observed at radius 270m and the minimum NDVI of 0.48 was observed at radius 90m. Since the NDVI values were obtained at few radii, we use linear interpolation to estimate individual NDVI values as continuous NDVI functions $X_i(s)$.  

\renewcommand{\thefigure}{4}
\begin{figure}[h!]
\centering
\includegraphics[scale=0.45]{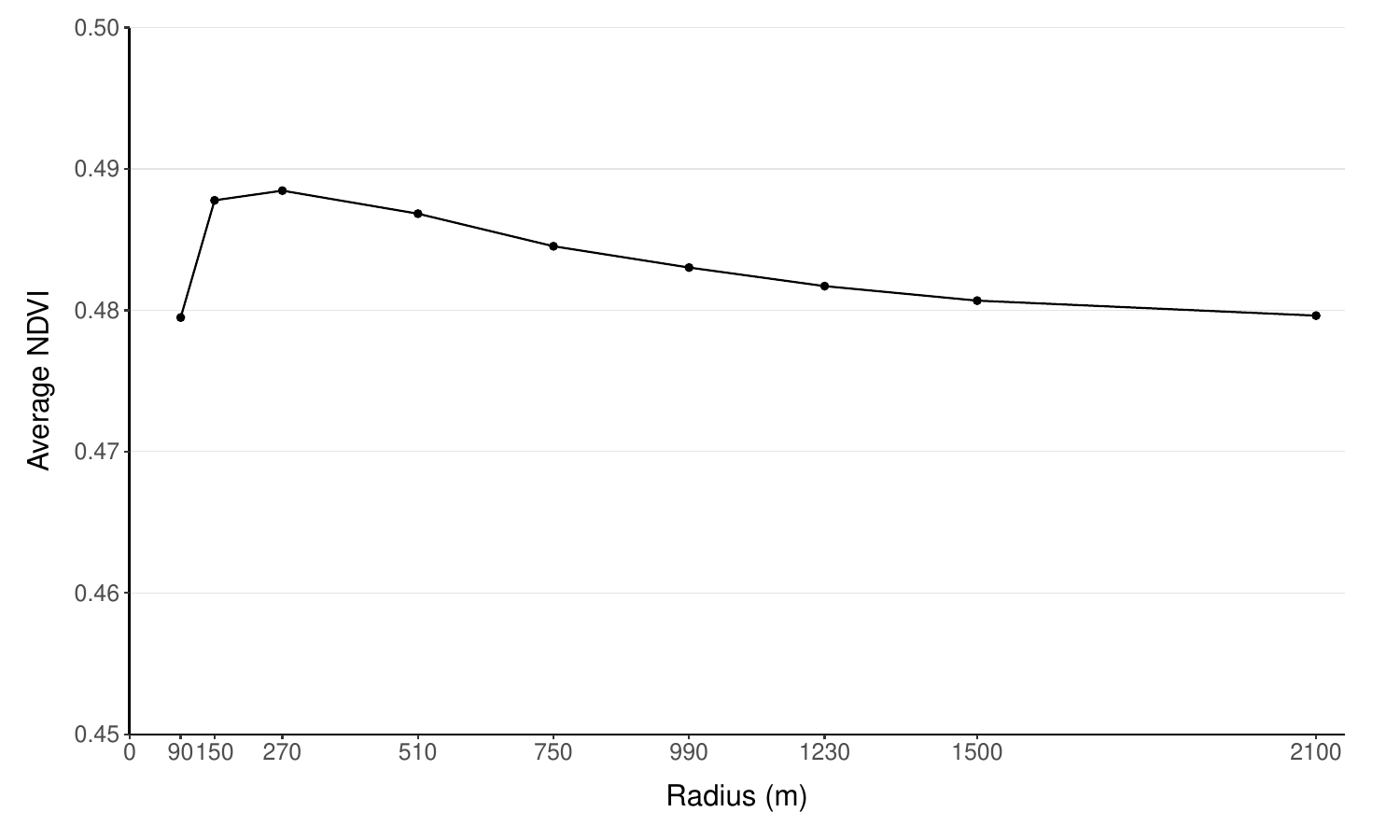}
\caption{The average NDVI values in concentric rings based on circular distance buffers at radii of 0m, 90m, 150m, 270m, 510m, 750m, 990m, 1230m, 1500m, and 2100m  for a total of 35,283 eligible participants in NHS.}
\label{Figure4}
\end{figure}

In a naive analysis,  a Cox regression model was fitted for each circular distance buffer separately. To apply our proposed method,  we used the two-stage approach described in Section \ref{Sec3.6}, where in the first stage, we fit the functional linear Cox model with the sparseness and smoothness penalties to identify non-null regions and find the buffer distance by the supremum of a set of non-null regions, and in the second stage, we re-fit the functional linear Cox model with the smoothness penalty within the buffer distance obtained from the first stage. In both the naive analysis and the analysis using our method, we stratified by age in months and additionally adjusted for potential risk factors including race, smoking status, pack-years of smoking, alcohol consumption, marital status, educational attainment, husband’s educational attainment, baseline Mental Health Inventory-5 score, and area-level characteristics such as Census tract population density, Census tract median income, and Census tract median home value. We set the B-spline basis $\{B_j(s): j=1, \ldots, 9\}$ with degree  3,  $M_n=5$, and inner knots at 150m, 270m, 510m, 990m, 1500m, and $\mathcal{S}$ as $[90, 2100]$m. The tuning parameters were selected based on BIC.

\begin{table}[h!]
\centering
\begin{tabular}{cccc}
\hline
Radius(\text{m}) & Hazard ratio & 95\% CI & p-value \\
\hline
$90$ & 0.952 & (0.928, 0.976) & 0.0001\\
$150$ & 0.950 & (0.926, 0.975) & $<0.0001$ \\
$270$ & 0.951 & (0.926, 0.976) & 0.0001 \\
$510$ & 0.955 & (0.930, 0.980) & 0.0005 \\
$750$ & 0.955 & (0.930, 0.980) & 0.0005 \\
$990$ & 0.956 & (0.932, 0.982) & 0.0008 \\
$1230$ & 0.958 & (0.933, 0.983) & 0.0011   \\
$1500$  & 0.958 & (0.934, 0.983) & 0.0011 \\
$2100$ & 0.959 & (0.935, 0.983) & 0.0011 \\
\hline
\end{tabular}
\caption{The estimated hazard ratios, confidence intervals, and p-value for 0.1 unit increase in the average NDVI within each of nine circular distance buffers (90m, 150m, 270m, 510m, 750m, 990m, 1230m, 1500m, and 2100m) based on separate stratified Cox regression models.}
\label{Table3}
\end{table}

Table~\ref{Table3} shows the results of naive analysis including the hazard ratios (HRs), confidence intervals, and p-value of 0.1 unit increase in the average NDVI within nine circular distance buffers, respectively, using the stratified Cox regression model. The NDVI decreased the risk of depression in all nine circular distance buffers, and the risk of depression was smallest at 150m circular distance buffer (HR: 0.950, 95\% CI: 0.926–0.975, p-value$<0.001$) and the risk of depression increased as the buffer distance increased. In the naive analysis, we were not able to estimate the effect of NDVI as a function of radius; therefore, it was not possible to identify the buffer distance. 

In our proposed model, the non-null regions of $[0, 510]$m were identified; thus, the buffer distance of 510m was chosen. Given the buffer distance of 510m, $\hat{\beta}^*(s)$ and its variance were estimated using the functional Cox model with the smoothness penalty. Figure~\ref{Figure5} shows the hazard ratio of 0.1 unit increase in the average NDVI at a radius in meters of $s$ given the 510m buffer distance, i.e., $\exp\left(0.1\hat{\beta}^*(s) \right)$ while controlling for the NDVI values at different radii and confounders. The estimated cumulative effect for 0.1 unit increase in the average NDVI within the 510m buffer distance $\left(\exp\left(\int_0^{510} 0.1\hat{\beta}^*(s) \right)\right)$ was 0.946 (95\% CI: 0.920–0.973, p-value=0.0001), which was comparable to the results based on the naive analysis using the 510m circular distance buffer (HR: 0.955, 95\% CI:  0.930–0.980, p-value=0.0005).

\renewcommand{\thefigure}{5}
\begin{figure}[h!]
\centering
\includegraphics[scale=0.6]{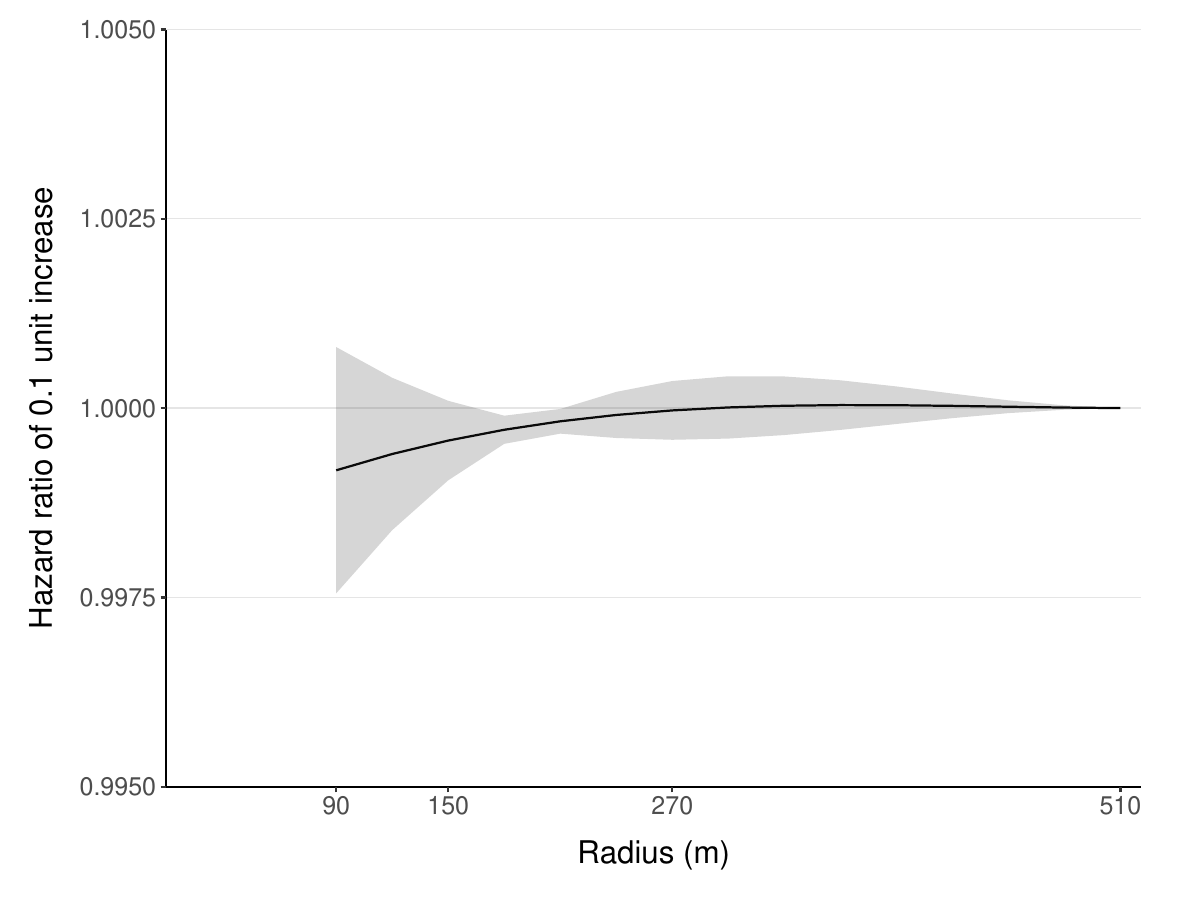}
\caption{The solid black line is the estimated hazard ratio of 0.1 unit increase in the average NDVI and the shaded area (gray) is its 95\% confidence interval as a function of radius within the buffer distance of 510m using the proposed two-stage method.}
\label{Figure5}
\end{figure}

 

%% file: sec9.tex
\section{Discussion}\label{sec:dis}

In our greenness and depression study, we investigated the risk of depression associated with the greenness of the neighborhood environment and identified the data-informed optimal buffer distance used to measure green space exposure. To overcome the limitations of traditional analysis,  which uses radial buffer distances selected a {\it priori}, we propose the two-stage functional Cox regression model, in which the optimal buffer distance is identified in the first stage, and given this optimal buffer distance, we estimate the functional association between incident depression and residential greenness exposure. Our analysis result using this proposed method suggests that the greenness within a 510m buffer distance from the residence had a protective effect on the risk of developing depression, and the protective effect was strongest at 90m and decreased as the distance from the residence increased. 

In our study based on the NHS, NDVI exposure was sparsely measured at only nine different circular distance buffers and we used the linear interpolated NDVI exposures to obtain regular and dense functional data. The identification of buffer distance may be sensitive to the choice of number and locations of knots in the B-spine basis functions; therefore, dense NDVI exposure may be required for future research. In addition, the statistical inference in the second stage does not take into account the variability induced by the identification of non-null regions.  The development of appropriate statistical models to account for post-selection inference in functional linear models is a topic for our future research. Furthermore, our proposed method is based on finite basis functions such as B-spline functions, and we applied the theoretical results in the functional linear Cox models under the reproducing kernel Hilbert space (RKHS) \citep{hao2021semiparametric} to the second stage in our proposed two-stage approach. It may be of interest to apply the RKHS framework to the selection problem in functional coefficients for future research. 

%% file: sec10.tex
\section*{Supplementary Materials}

The Supplementary Materials contain detailed proofs of all theorems, additional simulation results, and R code for the simulation studies and the data example.

\section*{Acknowledgements}
This project was supported by the National Research Foundation of Korea grant funded by the Korean government (MSIT) (No. 2022R1C1C1011806), the NIH/NIEHS grant R01ES026246, P30 ES00000, UM1  CA186107, the NIH/NHLBI grant R01 HL150119, and the DoD grant W81XWH2210030.

%% file: ref.bib
@article{banay2019greenness,
  title={Greenness and depression incidence among older women},
  author={Banay, Rachel F and James, Peter and Hart, Jaime E and Kubzansky, Laura D and Spiegelman, Donna and Okereke, Olivia I and Spengler, John D and Laden, Francine},
  journal={Environmental Health Perspectives},
  volume={127},
  number={2},
  pages={027001},
  year={2019}
}

@article{james2014effects,
  title={Effects of buffer size and shape on associations between the built environment and energy balance},
  author={James, Peter and Berrigan, David and Hart, Jaime E and Hipp, J Aaron and Hoehner, Christine M and Kerr, Jacqueline and Major, Jacqueline M and Oka, Masayoshi and Laden, Francine},
  journal={Health \& Place},
  volume={27},
  pages={162--170},
  year={2014},
  publisher={Elsevier}
}

@article{kwan2012uncertain,
  title={The uncertain geographic context problem},
  author={Kwan, Mei-Po},
  journal={Annals of the Association of American Geographers},
  volume={102},
  number={5},
  pages={958--968},
  year={2012},
  publisher={Taylor \& Francis}
}

@book{ramsay1997functional,
  title={Functional data analysis},
  author={Ramsay, James and Silverman, BW},
  year={1997},
  publisher={Springer}
}

@article{cardot2003spline,
  title={Spline estimators for the functional linear model},
  author={Cardot, Herv{\'e} and Ferraty, Fr{\'e}d{\'e}ric and Sarda, Pascal},
  journal={Statistica Sinica},
  pages={571--591},
  year={2003},
  publisher={JSTOR}
}

@article{gellar2015cox,
  title={Cox regression models with functional covariates for survival data},
  author={Gellar, Jonathan E and Colantuoni, Elizabeth and Needham, Dale M and Crainiceanu, Ciprian M},
  journal={Statistical Modelling},
  volume={15},
  number={3},
  pages={256--278},
  year={2015},
  publisher={SAGE Publications Sage India: New Delhi, India}
}

@article{qu2016optimal,
  title={Optimal estimation for the functional cox model},
  author={Qu, Simeng and Wang, Jane-Ling and Wang, Xiao and others},
  journal={The Annals of Statistics},
  volume={44},
  number={4},
  pages={1708--1738},
  year={2016},
  publisher={Institute of Mathematical Statistics}
}

@article{kong2018flcrm,
  title={FLCRM: Functional linear cox regression model},
  author={Kong, Dehan and Ibrahim, Joseph G and Lee, Eunjee and Zhu, Hongtu},
  journal={Biometrics},
  volume={74},
  number={1},
  pages={109--117},
  year={2018},
  publisher={Wiley Online Library}
}

@article{lin2017locally,
  title={Locally sparse estimator for functional linear regression models},
  author={Lin, Zhenhua and Cao, Jiguo and Wang, Liangliang and Wang, Haonan},
  journal={Journal of Computational and Graphical Statistics},
  volume={26},
  number={2},
  pages={306--318},
  year={2017},
  publisher={Taylor \& Francis}
}

@article{guan2020estimating,
  title={Estimating truncated functional linear models with a nested group bridge approach},
  author={Guan, Tianyu and Lin, Zhenhua and Cao, Jiguo},
  journal={Journal of Computational and Graphical Statistics},
  volume={29},
  number={3},
  pages={620--628},
  year={2020},
  publisher={Taylor \& Francis}
}

@article{hall2016truncated,
  title={Truncated linear models for functional data},
  author={Hall, Peter and Hooker, Giles},
  journal={Journal of the Royal Statistical Society: Series B (Statistical Methodology)},
  volume={78},
  number={3},
  pages={637--653},
  year={2016},
  publisher={Wiley Online Library}
}

@article{zou2005regularization,
  title={Regularization and variable selection via the elastic net},
  author={Zou, Hui and Hastie, Trevor},
  journal={Journal of the Royal Statistical Society: Series B (statistical methodology)},
  volume={67},
  number={2},
  pages={301--320},
  year={2005},
  publisher={Wiley Online Library}
}

@book{deboor2001spline,
  title={A practical guide to splines},
  author={De Boor, Carl},
  year={2001},
  publisher={Springer-Verlag New York}
}

@article{james2009functional,
  title={Functional linear regression that’s interpretable},
  author={James, Gareth M and Wang, Jing and Zhu, Ji and others},
  journal={The Annals of Statistics},
  volume={37},
  number={5A},
  pages={2083--2108},
  year={2009},
  publisher={Institute of Mathematical Statistics}
}

@article{wang2015functional,
  title={Functional sparsity: Global versus local},
  author={Wang, Haonan and Kai, Bo},
  journal={Statistica Sinica},
  pages={1337--1354},
  year={2015},
  publisher={JSTOR}
}

@article{huang2014group,
  title={Group selection in the Cox model with a diverging number of covariates},
  author={Huang, Jian and Liu, Li and Liu, Yanyan and Zhao, Xingqiu},
  journal={Statistica Sinica},
  pages={1787--1810},
  year={2014},
  publisher={JSTOR}
}

@article{huang2009group,
  title={A group bridge approach for variable selection},
  author={Huang, Jian and Ma, Shuange and Xie, Huiliang and Zhang, Cun-Hui},
  journal={Biometrika},
  volume={96},
  number={2},
  pages={339--355},
  year={2009},
  publisher={Oxford University Press}
}

@article{zhang2007adaptive,
  title={Adaptive Lasso for Cox's proportional hazards model},
  author={Zhang, Hao Helen and Lu, Wenbin},
  journal={Biometrika},
  volume={94},
  number={3},
  pages={691--703},
  year={2007},
  publisher={Oxford University Press}
}

@article{huang2004polynomial,
  title={Polynomial spline estimation and inference for varying coefficient models with longitudinal data},
  author={Huang, Jianhua Z and Wu, Colin O and Zhou, Lan},
  journal={Statistica Sinica},
  pages={763--788},
  year={2004},
  publisher={JSTOR}
}

@article{huang2010variable,
  title={Variable selection in nonparametric additive models},
  author={Huang, Jian and Horowitz, Joel L and Wei, Fengrong},
  journal={Annals of Statistics},
  volume={38},
  number={4},
  pages={2282},
  year={2010},
  publisher={NIH Public Access}
}

@article{marx1999generalized,
  title={Generalized linear regression on sampled signals and curves: a P-spline approach},
  author={Marx, Brian D and Eilers, Paul HC},
  journal={Technometrics},
  volume={41},
  number={1},
  pages={1--13},
  year={1999},
  publisher={Taylor \& Francis Group}
}

@article{guo2016spline,
  title={Spline-Lasso in high-dimensional linear regression},
  author={Guo, Jianhua and Hu, Jianchang and Jing, Bing-Yi and Zhang, Zhen},
  journal={Journal of the American Statistical Association},
  volume={111},
  number={513},
  pages={288--297},
  year={2016},
  publisher={Taylor \& Francis}
}

@article{baek2016distributed,
  title={Distributed lag models: examining associations between the built environment and health},
  author={Baek, Jonggyu and S{\'a}nchez, Brisa N and Berrocal, Veronica J and Sanchez-Vaznaugh, Emma V},
  journal={Epidemiology (Cambridge, Mass.)},
  volume={27},
  number={1},
  pages={116},
  year={2016},
  publisher={NIH Public Access}
}

@article{hao2021semiparametric,
  title={Semiparametric inference for the functional Cox model},
  author={Hao, Meiling and Liu, Kin-yat and Xu, Wei and Zhao, Xingqiu},
  journal={Journal of the American Statistical Association},
  volume={116},
  number={535},
  pages={1319--1329},
  year={2021},
  publisher={Taylor \& Francis}
}

@article{spiegelman2010approaches,
  title={Approaches to uncertainty in exposure assessment in environmental epidemiology},
  author={Spiegelman, Donna},
  journal={Annual Review of Public Health},
  volume={31},
  pages={149--163},
  year={2010},
  publisher={Annual Reviews}
}

@article{van2019environmental,
  title={Environmental exposures and depression: biological mechanisms and epidemiological evidence},
  author={van den Bosch, Matilda and Meyer-Lindenberg, Andreas},
  journal={Annual Review of Public Health},
  volume={40},
  pages={239--259},
  year={2019},
  publisher={Annual Reviews}
}

@article{sarkar2018residential,
  title={Residential greenness and prevalence of major depressive disorders: a cross-sectional, observational, associational study of 94879 adult UK Biobank participants},
  author={Sarkar, Chinmoy and Webster, Chris and Gallacher, John},
  journal={The Lancet Planetary Health},
  volume={2},
  number={4},
  pages={e162--e173},
  year={2018},
  publisher={Elsevier}
}

@article{pun2018association,
  title={Association of neighborhood greenness with self-perceived stress, depression and anxiety symptoms in older US adults},
  author={Pun, Vivian C and Manjourides, Justin and Suh, Helen H},
  journal={Environmental Health},
  volume={17},
  pages={1--11},
  year={2018},
  publisher={Springer}
}

@article{liu2019neighbourhood,
  title={Neighbourhood greenness and mental wellbeing in Guangzhou, China: What are the pathways?},
  author={Liu, Ye and Wang, Ruoyu and Grekousis, George and Liu, Yuqi and Yuan, Yuan and Li, Zhigang},
  journal={Landscape and Urban Planning},
  volume={190},
  pages={103602},
  year={2019},
  publisher={Elsevier}
}

@article{james2018global,
  title={Global, regional, and national incidence, prevalence, and years lived with disability for 354 diseases and injuries for 195 countries and territories, 1990--2017: a systematic analysis for the Global Burden of Disease Study 2017},
  author={James, Spencer L and Abate, Degu and Abate, Kalkidan Hassen and Abay, Solomon M and Abbafati, Cristiana and Abbasi, Nooshin and Abbastabar, Hedayat and Abd-Allah, Foad and Abdela, Jemal and Abdelalim, Ahmed and others},
  journal={The Lancet},
  volume={392},
  number={10159},
  pages={1789--1858},
  year={2018},
  publisher={Elsevier}
}

@techreport{world2017depression,
  title={Depression and other common mental disorders: global health estimates},
  author={{World Health Organization}},
  year={2017},
  institution={World Health Organization}
}

@article{yuan2010reproducing,
  title={A REPRODUCING KERNEL HILBERT SPACE APPROACH TO FUNCTIONAL LINEAR REGRESSION},
  author={Yuan, Ming and Cai, T Tony},
  journal={The Annals of Statistics},
  volume={38},
  number={6},
  pages={3412--3444},
  year={2010}
}

@article{rhew2011validation,
  title={Validation of the normalized difference vegetation index as a measure of neighborhood greenness},
  author={Rhew, Isaac C and Vander Stoep, Ann and Kearney, Anne and Smith, Nicholas L and Dunbar, Matthew D},
  journal={Annals of Epidemiology},
  volume={21},
  number={12},
  pages={946--952},
  year={2011},
  publisher={Elsevier}
}

@book{bishop2006pattern,
  title={Pattern recognition and machine learning},
  author={Bishop, Christopher M and Nasrabadi, Nasser M},
  volume={4},
  number={4},
  year={2006},
  publisher={Springer}
}
